\documentclass[prl, aps, twocolumn, superscriptaddress, nofootinbib, tightenlines, nobibnotes, showpacs, 10pt]{revtex4-1}

\usepackage{natbib}
\usepackage{slashed}
\usepackage{graphicx}
\usepackage{subfigure}
\usepackage[usenames, dvipsnames]{color}
\usepackage{graphics}
\usepackage{hyperref}
\usepackage{bm}
\usepackage{comment}
\usepackage{amsmath}
\usepackage{color}\label{key}
\usepackage{amsfonts}
\hypersetup{backref,
colorlinks=true,
linkcolor=blue,
linktoc=page,
citecolor=blue,
urlcolor=blue}

\everymath{\displaystyle}

\begin{document}

\title{Information compression at the turbulent-phase transition in cold atom gases}

\author{R. Giampaoli}
\affiliation{\it Instituto de Plasmas e Fus\~{a}o Nuclear, Instituto Superior T\'{e}cnico,
Universidade de Lisboa, 1049-001 Lisbon, Portugal}
\affiliation{\it Laboratoire Kastler Brossel, Sorbonne Universit\'{e}, CNRS, ENS-PSL Research University, Coll\`{e}ge de France, Paris, 75005, France}

\author{J. L. Figueiredo}
\affiliation{\it Instituto de Plasmas e Fus\~{a}o Nuclear, Instituto Superior T\'{e}cnico,
Universidade de Lisboa, 1049-001 Lisbon, Portugal}

\author{J. D. Rodrigues}
\affiliation{\it Instituto de Plasmas e Fus\~{a}o Nuclear, Instituto Superior T\'{e}cnico,
Universidade de Lisboa, 1049-001 Lisbon, Portugal}

\author{J. A. Rodrigues}
\affiliation{\it Instituto de Plasmas e Fus\~{a}o Nuclear, Instituto Superior T\'{e}cnico,
Universidade de Lisboa, 1049-001 Lisbon, Portugal}
\affiliation{\it Departamento de F\'{i}sica, Universidade do Algarve, Campus de Gambelas, 8005-139 Faro, Portugal}

\author{H. Ter\c{c}as}
\affiliation{\it Instituto de Plasmas e Fus\~{a}o Nuclear, Instituto Superior T\'{e}cnico,
Universidade de Lisboa, 1049-001 Lisbon, Portugal}

\author{J. T. Mendon\c{c}a}
\affiliation{\it Instituto de Plasmas e Fus\~{a}o Nuclear, Instituto Superior T\'{e}cnico,
Universidade de Lisboa, 1049-001 Lisbon, Portugal}

\begin{abstract}

The statistical properties of physical systems in thermal equilibrium are blatantly different from their far-from-equilibrium counterparts. In the latter, fluctuations often dominate the dynamics and might cluster in ordered patterns in the form of dissipative coherent structures. Here, we study the transition of a cold atomic cloud, driven close to a sharp electronic resonance, from a stable to a turbulent phase. From the atomic density distribution --- measured using a spatially-resolved pump-probe technique --- we have computed the Shannon entropy on two different basis sets. Information compression, corresponding to a minimum in the Shannon entropy, has been observed at criticality, where the system fluctuations organize into high-order (low-entropy) patterns. Being independent of the representation used, this feature is a property shared by a vast class of physical systems undergoing phase transitions. 
\end{abstract}

\maketitle 

%
%
\textit{Introduction.}---The macroscopic properties of any physical system in equilibrium can be completely described in terms of free-energy landscapes, independently of the system microscopic details \cite{GibbsEnergyLandscapes, jaynes1980minimum}. Far from thermodynamic equilibrium, however, the same results do not apply \cite{jaynes1986predictive, hinrichsen2006non, Joao_Jitter} and even how to rigorously define thermodynamic potentials has been controversial for many years and it is still a matter of debate \cite{jaynes1980minimum, parrondo2015thermodynamics}. In far from-equilibrium conditions, spatiotemporal dynamics are dominated by fluctuations, which get amplified and often lead to the formation of coherent dissipative structures \cite{prigogine1975dissipative, prigogine1978time}. Some of these phenomena, where order spontaneously emerge from stochastic fluctuations, can be described with a formalism analogous to the one used for equilibrium phase transitions, a striking example being the analogy of a laser threshold region and a second order phase transition \cite{degiorgio1970analogy, Joao_Jitter}. Moreover, when phase transitions involve loss of global symmetry, the main statistical properties can often be captured by a single order parameter, as in the case of stable-to-turbulence transitions \cite{egolf2017mean, egolf2020nonlinear}. In general, however, despite the existence of several theoretical models, when it comes to phase transitions of non-equilibrium systems evidences of universality are still very poor \cite{hinrichsen2006non}.\par
In recent years, a possible extension of the methods used in thermodynamic equilibrium to far-from-equilibrium scenarios has been proposed by resorting to information theory \cite{brillouin1953negentropy, jaynes1980minimum, jaynes1986predictive, parrondo2015thermodynamics}. The description of physical quantities in such terms requires a much more limited set of assumptions and finds applications in different fields, such as astrophysics \cite{heavens_2020}. Information (Shannon) entropy is a key quantity in the information-theory analysis of phase transitions, playing the role of its thermodynamic (Boltzmann) counterpart. In a phase transition, the state of a system may be described by some generic field $\Phi\equiv \Phi(\mathbf{r},t;\delta)$, where $\delta$ is the stress parameter (often, but not always, the thermodynamic temperature). Given a complete and orthogonal basis for the space of square-integrable functions, the field can be uniquely identified by the projections onto the basis elements. The link between the thermodynamic and information theories is provided by the expansion coefficients: they can be viewed both as the spectral energy content of the system (that is spread over several physical modes) and, at the same time, as the probability distribution which encodes the information of each mode \cite{parrondo2015thermodynamics, sowinski2017information}. A phase transition can then be viewed as a change in the group symmetries of the field when the stress parameter attains some critical value $\delta_c$. Typically, the transition mechanism can be described by a single (or few) order parameter $\psi$, which breaks the original symmetry by assuming a non-zero value \cite{landau2013statistical}.
\begin{figure}[!t]
\centering
\includegraphics[width=3.3in]{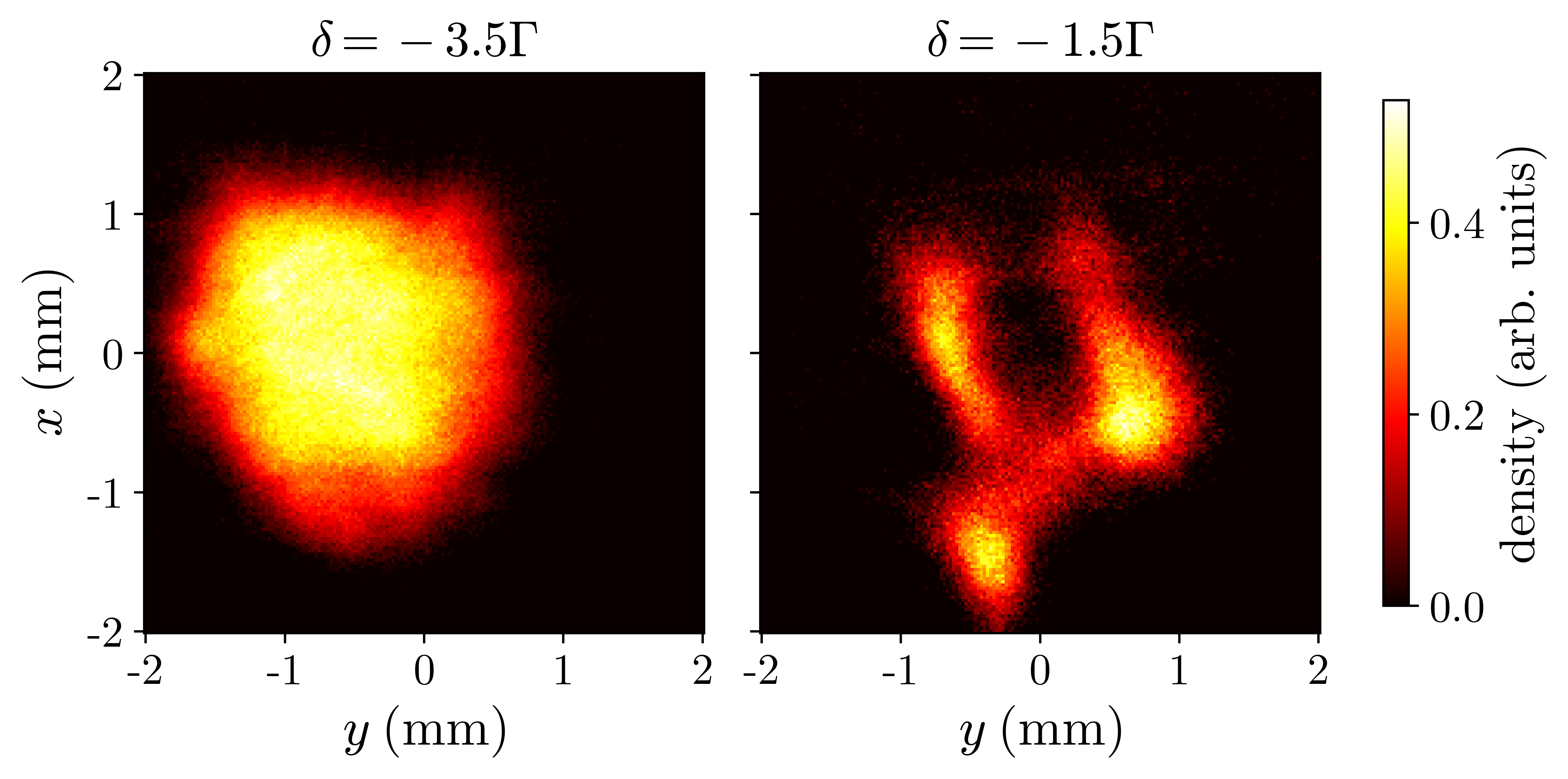}
\caption{
Snapshots of a single realization of the atomic cloud density distribution in the stable phase ($\delta = -3.5\Gamma$) and in the turbulent phase ($\delta=-1.5\Gamma$). 
The cold atom cloud in the turbulent regime is highly inhomogeneous and is charactered by strong spatiotemporal dynamics which feature the emergence of quasi-coherent structures.
The images have been retrieved with the spatially-resolved pump-probe technique described in \cite{giampaoli2021photon}.
}
\label{fig:snapshots} 
\end{figure}
%
\begin{figure*}[!t]
\centering
\includegraphics[width=5.8in]{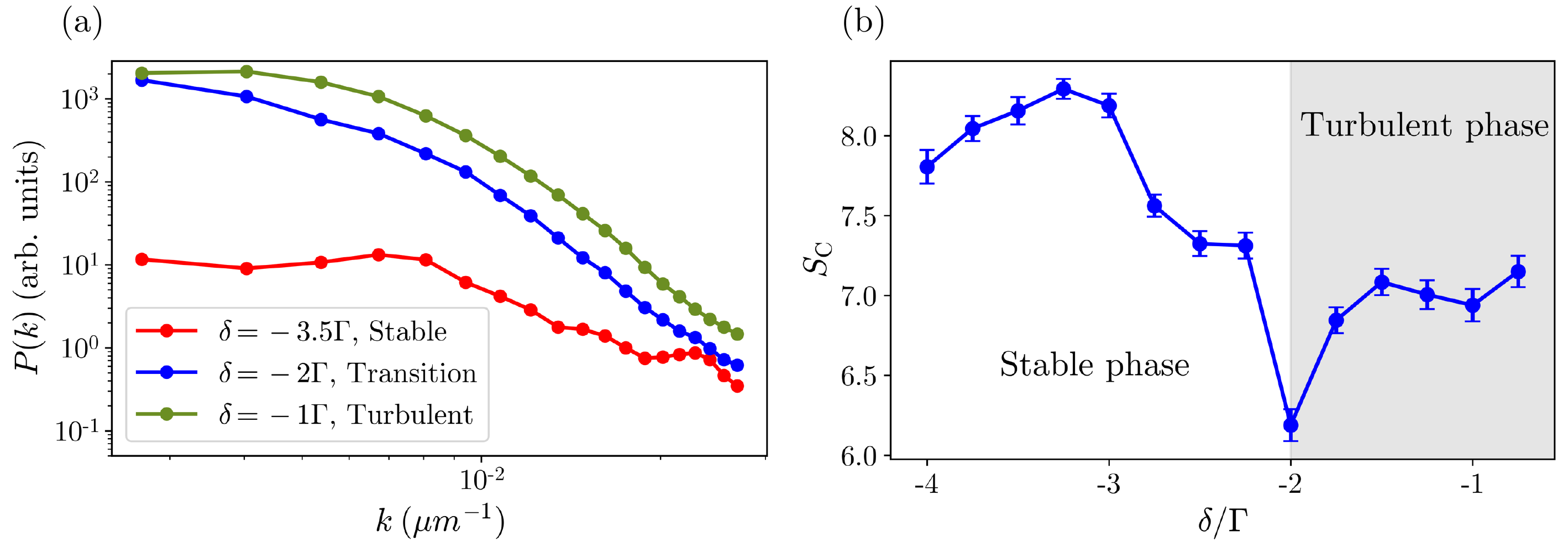}
\caption{Panel (a): radial component of the 2D average power spectrum in the stable and turbulent regimes and at criticality. In terms of magnitude, the turbulence phase is characterized by much larger fluctuations, when comparing to the stable phase. Panel (b): configurational entropy computed from the spatial power spectrum (see Eq. \eqref{Eq:Configurational_entropy}) as a function of the stress parameter \( \delta\).
In the stable regime, the flatter power spectrum is reflected onto higher values of entropy.
At $\delta =-2\Gamma$, the highly compact spectral distribution is at the origin of the entropy dip.
}
\label{fig:Configurational_entropy} 
\end{figure*}
\par
In this Letter, we report the observation of information compression at the stable-turbulent phase transition in a cold atom cloud, occurring when the cooling laser frequency is set near the electronic resonance \cite{giampaoli2021photon, gaudesius_2022}. Measures have been performed on a cold rubidium gas by directly probing various statistically independent realizations of the atomic density distribution (the field \(\Phi\)) whose statistical properties are controlled by the cooling laser detuning \(\delta\), which acts as stress parameter. By varying \(\delta\), the system goes from a stable-symmetric (uniform) phase to a turbulent one, where the global symmetry is lost and quasi-coherent structures emerge. At criticality the system spontaneously organizes, showing both a long-range local order and formation of oscillating global patterns. We computed the Shannon entropy on two different basis for the multiple realizations of the atomic density distribution and, at the critical point, an entropy minimum---associated with information compression---is observed. Remarkably, this minimum is independent of the representation used, which points out the transition as a state of maximum organization and emerges as a potentially universal property of a large class of phase transitions. The latter property is verified with the help of a symmetry-breaking mechanism contained in the Landau-Ginzburg theory. 

%
%
\par
\textit{Turbulent-phase transition in a cold rubidium gas.}--- Experiments have been performed on a magneto-optical trap (MOT) \cite{raab1987trapping}, where around 10\textsuperscript{9} $^{85}$Rb atoms are cooled and trapped at approximately $200\;\!\mu$K \cite{JoaoEqState, rodrigues2016collective}. A spatially-resolved pump-probe diagnostic allows to directly access the atomic density distribution along a thin section of the atomic cloud. As a result, the collected images can be safely interpreted as two-dimensional (2D) atomic density maps \cite{giampaoli2021photon}. The experiments have been carried out by keeping the magnetic field approximately constant (\(  \nabla B = 10 \; \mathrm{G/cm}\)) and ranging the cooling laser detuning \(\delta\) from \( -4 \Gamma \) to \( - 0.75 \Gamma \), with $\Gamma$ denoting the transition linewidth. For each \( \delta\), 100 measurements have been performed, effectively probing different system realizations. Each measurement consists of a loading step, during which the atomic gas is trapped and cooled until it forms a stationary cloud, followed by the MOT unloading, after which the pump-probe sequence is executed.
When the frequency of the cooling lasers is brought close to resonance, the cold atom cloud passes through a sharp transition from a stable, spatially uniform, phase to a turbulent phase \cite{giampaoli2021photon}. Fig. \ref{fig:snapshots} shows the density profiles of the atomic cloud in both phases. 
The turbulent regime is characterized by strong spatiotemporal fluctuations which develop as the cold atom gas is continuously cooled and trapped in the range \( \delta \in [-2 \Gamma, -0.75 \Gamma] \). 
The stable-turbulent transition is marked by an abrupt increase in the power of density fluctuations and by a peak in the fluctuation correlation length at the transition onset \( \delta = -2 \Gamma \). The turbulent dynamics originate from a fluid-dynamic instability, known as photon-bubble instability \cite{mendoncca2012photon}, which stems from the strong coupling of the atomic fluid with diffusive radiation. Photon bubbles leave a clear signature in the atomic fluctuation density in the form of quasi-coherent structures. As we shall see, an information-theory approach allows to capture and encapsulate the system dynamics in a single quantity---the Shannon entropy---thus providing an effective description of the transition of the cold atom system from the stable to the turbulent phase.
%
%
\par
\textit{Information entropy.}---Information theory \cite{shannonTheoryComm1948} provides the tools to measure the quantity of information resulting from the observation of an event.
The information content $I_j$ \cite{hartley1928transmission} is defined in such a way that unlikely events carry more information and it is computed as $I_j \equiv \mathrm{log}(1/p_j) = - \mathrm{log}(p_j)$, where  $p_j$ is the probability of the $j-$th event, and normalization requires $\sum\nolimits_j  p_j = 1$. 
Shannon entropy $S$ is then defined as the average information content per event: $S [\{p_j\}] = - \sum\nolimits_{j} p_j \mathrm{log} (p_j)$.
The maximum entropy corresponds to a flat distribution, in which no outcome is favored: each observation produces, on average, a high amount of information.
Hence, by comparing the value of \(S\) of the actual probability distribution to its maximum value attained for a flat distribution over the same ensemble, we get a quantitative measure of information compression. Highly compressed ensembles are characterized by clustered probability distributions which result in low entropy values. \par 
In order to interpret critical phenomena, we apply the same information-theory description to the field \(\Phi(\mathbf{r},t;\delta)\)---the density maps in our case---by decomposing it onto its (time-independent) basis elements.
Independently of the basis choice, the meaning of low entropy values is the same: a low number of highly probable modes dominates the system dynamics, meaning that less information, in terms of contributing momentum modes, is needed to characterize the system, resulting in field's compression in information space.
\par
The Shannon entropy defined from the field expansion onto Fourier modes is called configurational entropy, first introduced in Ref. \cite{gleiser2012entropic} as a measure for localized energy configurations.
Configurational entropy, $S_\mathrm{C}$, is defined in terms of the modal fraction $f(\mathbf{k})=P(\mathbf{k})/ \sum\nolimits_{\mathbf{k}} P(\mathbf{k})$ as
\begin{equation} \label{Eq:Configurational_entropy}
	S_\mathrm{C} = - \sum_{\mathbf{k}}    f(\mathbf{k}) \mathrm{log} \big[ f(\mathbf{k}) \big], \end{equation}  
where $P(\mathbf{k}) = \int \mathrm{d\mathbf{r}} \  e^{-i{\bf k} \cdot{\bf r}} \langle \Tilde{\Phi}(\mathbf{r})\Tilde{\Phi}(0)\rangle $ is the 2D power spectrum, $\tilde \Phi(\mathbf{r}) = \Phi(\mathbf{r}) - \langle \Phi(\mathbf{r}) \rangle $ is 2D atomic density fluctuation and ${\bf k}=(k_x,k_y)$ the wavevector.
Here, the symbol $\langle .\rangle$ means averaging over all experimental realizations corresponding to the same stress parameter $\delta$ \cite{giampaoli2021photon}.
The left panel of Fig. \ref{fig:Configurational_entropy} shows the power spectra in the stable and unstable regimes, as well as at the critical point, $\delta\equiv \delta_c\simeq-2\Gamma$. \par
The configurational entropy is also a measure of a system spatial complexity, being able to pinpoint the formation of coherent structures: low entropy configurations are associated to the presence of (local) order. As in Shannon entropy, configurational entropy is maximum for a flat spectrum: \(S_\mathrm{C} =  \mathrm{log}(M)\), where \( M \) is the total number of modes. 
Information compression can then be regarded as a departure from a flat distribution, resulting from the emergence of local structures and being described by low entropy values.
\begin{figure*}[!ht]
\centering
\includegraphics[scale=0.55]{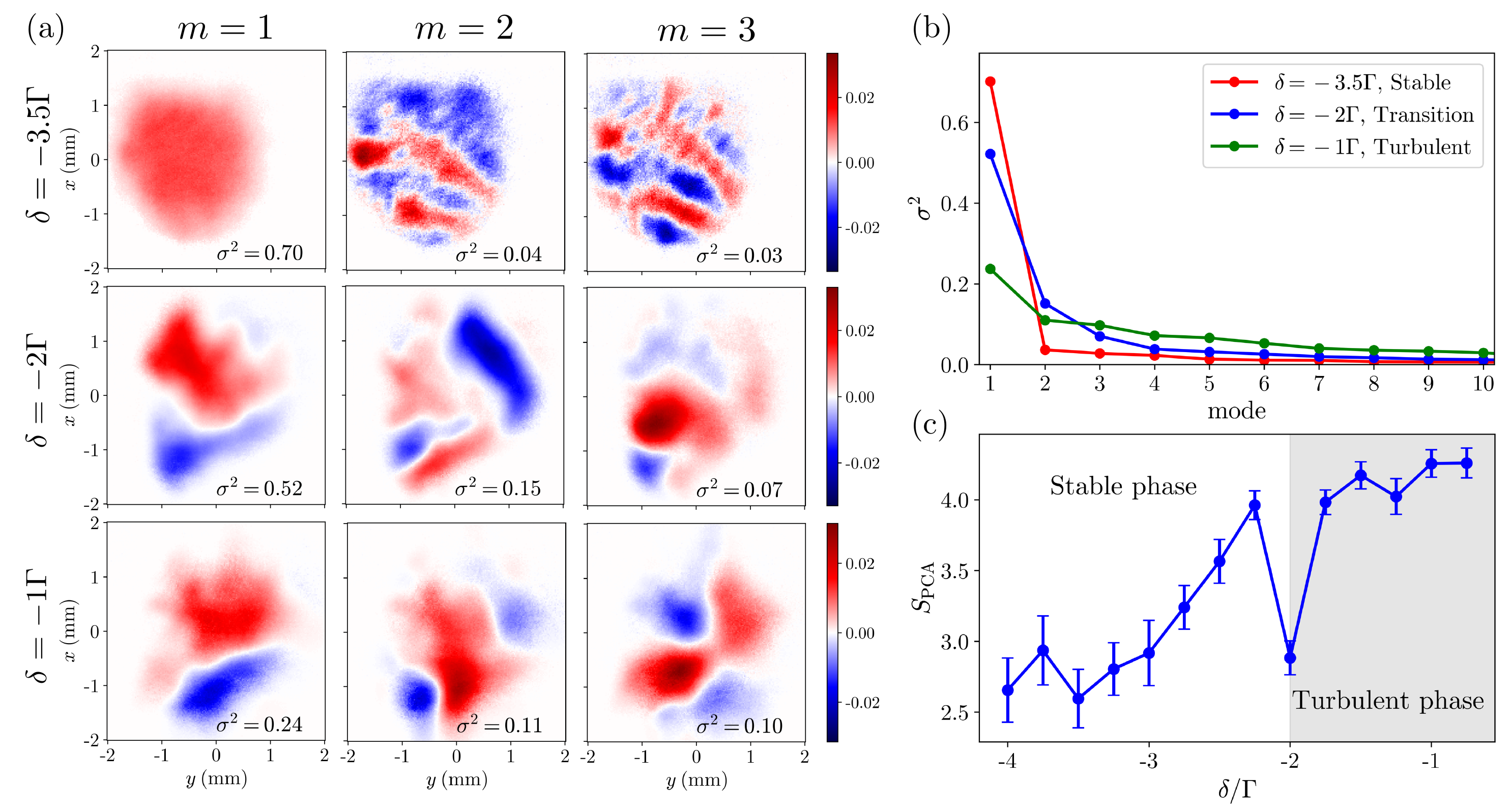}
\caption{
Panel (a): first three PCA modes in the stable regime (first row), at criticality (second row) and in the turbulent regime (third row); the color map represents density fluctuations. Panel (b): modal variance in the same regimes. Panel (c): entropy computed from the modal variance as a function of the detuning of the trapping lasers which drive the cold atom cloud.
Low values of entropy descend from a small amount of highly-representative (high variance) modes which describe the system at a given \(\delta \).
}
\label{fig:PCA_entropy} 
\end{figure*}
%
\par
Besides a Fourier spectral decomposition of the density fluctuation power, a study performed through a Principal Component Analysis (PCA) offers a different way to link information compression to the atomic density distribution. PCA is a model-free approach which provides a lower-dimensional representation of a given data set by writing it as a linear combination of statistically uncorrelated normal modes (principal components) \cite{bishop2006pattern, brunton2022data}. By carrying out a principal component analysis on the whole set of atomic density maps, we can portray each individual frame \( \Phi_i (x,y)\) as the sum of an average map and a linear combination of orthogonal principal component modes \(\mathcal{U} _m (x,y)\) \cite{rodrigues2022quasi},
%
\begin{gather} \label{Eq:PCA_definitions}
 \Tilde{\Phi}_i (x,y) = \sum_{m=1}^M  \lambda _{m,i} \, \mathcal{U} _m (x,y) \, ,  \\
\langle \mathcal{U}_m, \mathcal{U}_\ell\rangle  = \mathrm{Tr}(\mathcal{U} _m ^{\dagger} \mathcal{U} _\ell) = \delta _{m\ell} \, .
\end{gather}  
Above, the additional dependence on the stress parameter $\delta$ was omitted. 
The coefficients $\lambda_{m,j} \equiv \text{Tr}(\mathcal{U}_m^\dag \Tilde{\Phi}_j)$ are the projections onto the modes and we further define the mode variances as \( \sigma _m^2 = \langle \lambda_{m,i}^2\rangle \), denoting how much a given mode \( \mathcal{U} _m\) contributes to representing the ensemble. The number of principal components is given by the number of degrees of freedom of each realization (here, it corresponds to the number of pixels in the CCD), and we order them from the most to the least representative:
\begin{equation} \label{Eq:variance}
\sigma _1 ^2 \geq  \sigma _2 ^2 \geq  ... \geq  \sigma _M ^2 \geq 0, \; \; \; \sum_{m=1}^M \sigma _m ^2 = 1 .
\end{equation}  
The advantage of the analysis in terms of principal components is that, out of $M\sim32000$ modes, only a few are of relevance to characterize the system dynamics. That is expressed by having $\sigma^2_m \sim 0$ for $m>\overline{m}$, with $\overline{m}\ll M$ [see Fig.~\ref{fig:PCA_entropy} (b)], which lowers the dimensionality of the data set significantly. The first three PCA modes at three different values of the stress parameter---in the stable regime (\( \delta = -3 \Gamma\)), at criticality ($\delta = \delta_c$), and in the turbulent regime (\( \delta = -1 \Gamma\))---are depicted in Fig. \ref{fig:PCA_entropy}. Contrary to Fourier modes, which are independent of the data set, principal components are retrieved directly from the data set itself: Fourier and PCA modes are representative of two different classes of basis decomposition, the former being an example of ``universal" basis and, the latter, of ``tailored" basis. 
The PCA decomposition allows to highlight the similarities of the system dynamics along the whole range of $\delta$. 
Within the same regime (stable or turbulent), density fluctuations are described by analogous sets of principal components. 
Far in the stable region, the majority of the fluctuation power, captured in the first mode, is due to oscillations of the total number of atoms. 
Conversely, a limited but larger set of modes is necessary to characterize the turbulent regime. This is a feature reminiscent of low-dimensional chaos \cite{brandstater_1983, hu_2000, ott_2008, cestnik_2022}. 
\par
In complete analogy with the configurational entropy, information entropy based on the principal component decomposition can then be defined from the modal variances as
\begin{equation} \label{Eq:PCA_entropy}
S_\mathrm{PCA} = - \sum_{m} \sigma _m ^2 \mathrm{log} (\sigma _m ^2 )  \, .
\end{equation}  
It follows that high data compressibility descends from a small amount of oscillating global patterns whose linear combination is able to describe the main system dynamics. 
%
%
\par
\textit{Entropy signature at criticality.}---We shall now focus our attention on the stress parameter dependence of \(S_\mathrm{C}\) and \(S_\mathrm{PCA}\), interpreting the two curves in terms of spatial complexity and of the resulting modal distribution, i.e. the power spectrum and principal component variance.
In this regard, we should not overpass that the difference in the total amount of energy stored in density fluctuations in the stable and in the turbulent regimes is utterly irrelevant when it comes to entropy computation: what matters is just how energy is distributed among the modes. In the right panel of Fig. \ref{fig:Configurational_entropy}, we observe that, as \(\delta\) is brought from the stable to the turbulent region, the configurational entropy is reduced. 
In the stable regime, high entropy values characterize the highly symmetric, spatially uniform, physical system. The power spectrum is essentially flat as a consequence of the fact that the stable regime is dominated by uncorrelated fluctuations (white noise). Conversely, in the turbulent phase, the power spectrum is dominated by long wavelength modes, which, in the real space, witness the formation of randomly distributed large-scale
coherent structures. In other words, local order implies low values of configurational entropy.
Concerning \(S_\mathrm{PCA}\), as depicted in panel (c) of Fig. \ref{fig:PCA_entropy}, we observe the opposite dependence on $\delta$: low entropy values characterize the stable regime whereas high values are associated to the turbulent phase. These high values stem from the increased spatial complexity of the atomic cloud in the turbulent regime, where a larger number of principal components is necessary to represent the main system dynamics.
\par
Remarkably, however, both \(S_\mathrm{PCA}\) and \(S_\mathrm{C}\) show a narrow dip at the turbulence edge, $\delta = \delta_c$.
As the system moves from one phase to another, it passes through an intermediate step where it reorganizes:
the critical point is characterized by a divergence in the correlation length---analogous to critical opalescence \cite{walsh_2019, kundu_2020}---in a configuration where both local and global order coexist. 	
This highly structured state requires a low amount of information to be described, thus leading to a maximum in information compression.
This quantity is not only a measure of our knowledge of the system, but it is also a measure of complexity and order of a given spatial configuration, thus bearing information about the system physical properties.
\par 
Furthermore, theoretical and numerical studies have shown a configurational entropy dip at the onset of continuous equilibrium phase transitions \cite{gleiser2015information, sowinski2017information}. Here, the same feature has been experimentally observed for the first time for an out-of-equilibrium phase transition, thus raising Shannon entropy as a good candidate to interpret critical phenomena in light of information theory.
%
\par
\begin{figure}[!t]
\centering
\includegraphics[scale=0.55]{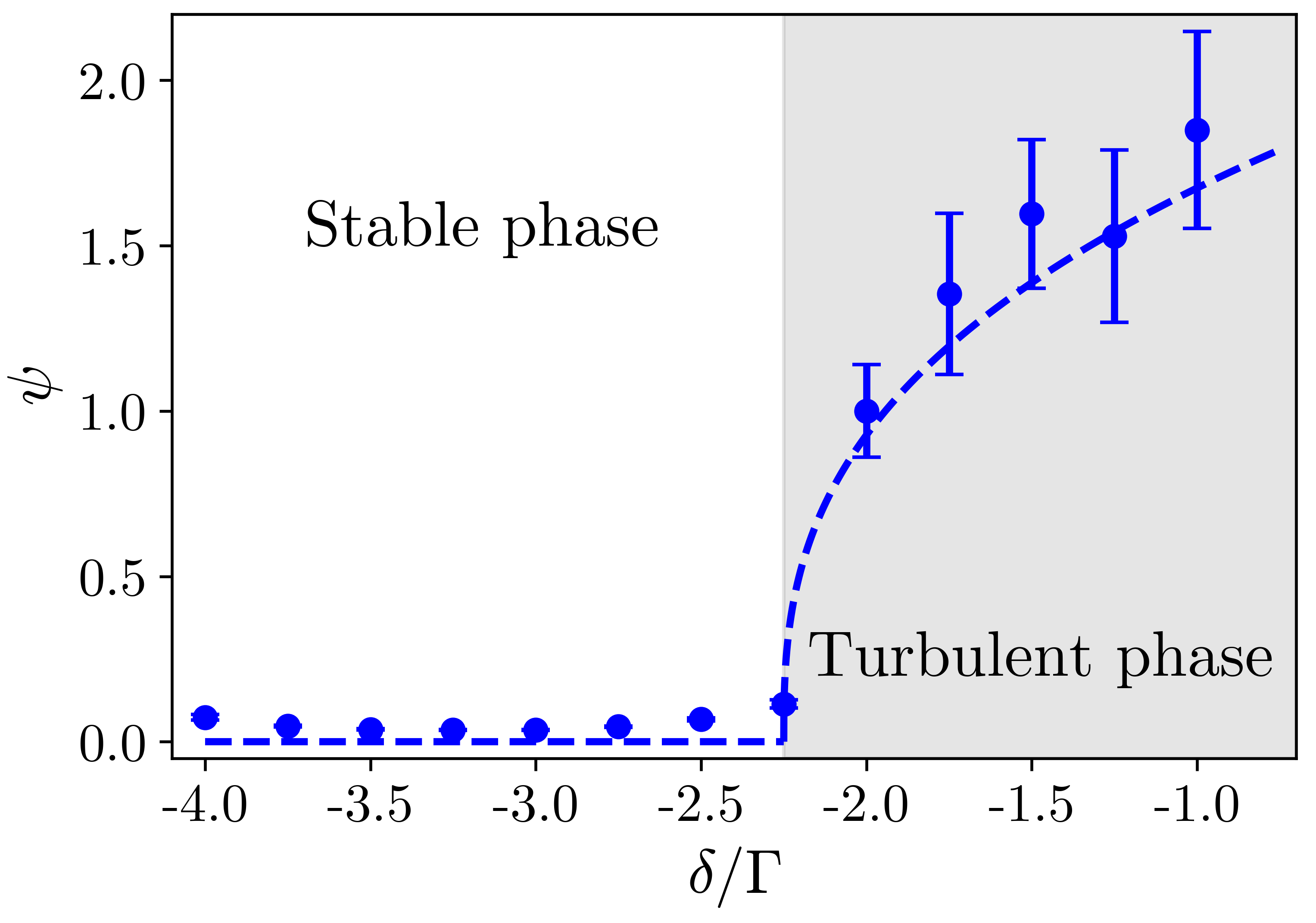}
\caption{Critical behavior of the total fluctuation power $\psi$ as a function of the stress parameter \(\delta \). The dots corresponds to the experimental data, while the dashed line is a fitting of the form $\psi=c(\delta-\delta_c)^{\beta}$, providing $\beta=0.372\pm 0.003$, and $c=1.54 \pm 0.15$. }
\label{fig:critical} 
\end{figure}

The universal character of the phase transition can be put in evidence within the Ginzburg-Landau framework. If we define the total fluctuation power $\psi= \sum\nolimits_{\mathbf k} P({\bf k}) $ as a tentative order parameter, the free energy can be written as 
\begin{equation}
\mathcal{F}[\psi]=\mathcal{F}_0 +a (\delta_c-\delta)  \psi^2+ \frac{b}{2} \psi^4,
\label{eq_free_energy}
\end{equation}
with $\mathcal{F}_0$ being the free energy deep in the stable phase, and $a$ and $b$ some positive constants. Minimization of Eq.~\eqref{eq_free_energy} yields the critical behavior $\psi=\left[(a/b) (\delta-\delta_c)\right]^{1/2}$ for $\delta > \delta_c$ and $\psi = 0$ otherwise. In Fig.~\ref{fig:critical}, we fit the experimental data to a test function of the form $\psi \sim \vert \delta-\delta_c \vert^\beta$, and observe that the critical parameter scales with the critical exponent $\beta^{\rm (exp.)}\simeq 0.372\pm 0.003$. The experimental value of the exponent must be compared with that given by the mean-field approximation, $\beta^{\rm (MF)}  =1/2$, which deviates from $\beta^{\rm (exp.)}$ due to large fluctuations taking place at the critical point. Experimental measurements of $\beta$ in other physical systems are in good agreement with that extracted from Fig.~\ref{fig:critical} \cite{criticalExp,criticalExp2}, which suggests that the turbulent phase transition observed here is well described in terms of the total power and falls into the universality class of second order phase transitions.
%


\par
\textit{Conclusions.}---Non-equilibrium phase transitions in driven-dissipative systems are still far from being completely understood and theories lack of experimental evidences.
In this manuscript we have reported the observation of a maximum in information compression at the critical point of a non-equilibrium phase transition---this being the transition of a cold atom cloud from a stable to a turbulent phase.
Information compression at criticality is witnessed by a dip in information entropy which has been computed in two different ways by projecting the atomic density fluctuations on two distinct basis sets, the Fourier modes and the principal components. 
The two modal decompositions are associated in a complementary way to system complexity, one highlighting local order and the other the presence of fluctuating global patterns. 
Notably, information compression at criticality is independent of the representation used.
\par
Our work is in keeping with the long-lasting line of research that aims to unify statistical mechanics with information theory, and brings a relevant contribution to the description of far-from-equilibrium physical systems. 
The experimental evidences we have provided show that information entropy, built on either of two different modal decompositions, is capable to pinpoint the transition of a cold atom cloud from a stable to an unstable phase. If shared by a vast class of physical systems, this feature could raise information entropy to a fundamental quantity to develop a unifying phase transition theory.

\par
\textit{Acknowledgments.}---We are grateful for the helpful discussion with Pedro Cosme during the elaboration of this work. R.G., J.L.F and H.T. acknowledge Funda\c{c}\~{a}o da Ci\^{e}ncia e a Tecnologia (FCT-Portugal) through the Grants No. PD/BD/135211/2017, UI/BD/151557/2021, and through Contract No. CEECIND/00401/2018 and Project No. PTDC/FIS-OUT/3882/2020, respectively. We further acknowledge funding from the European Union’s Horizon 2020 Research and Innovation programme under grant agreement no. 820392 (PhoQuS).

\bigskip

\bibliographystyle{apsrev4-1}
\bibliography{main_v6.bib}

\end{document}